# Can theory-driven learning analytics dashboard enhance human-AI collaboration in writing learning? Insights from an empirical experiment

Angxuan Chen[1], Jingjing Lian[2], Xinran Kuang[1], Jiyou Jia[1,*]

1. Department of Educational Technology, School of Education, Peking University, Beijing 100871, China
2. School of Humanities, Beijing University of Posts and Telecommunications, Beijing, 102206, China

* Corresponding Author: Jiyou Jia

Email: jjy@pku.edu.cn

**Abstract:** The integration of Generative AI (GenAI) into education has raised concerns about over-reliance and superficial learning, particularly in writing tasks in higher education. This study explores whether a theory-driven learning analytics dashboard (LAD) can enhance human-AI collaboration in the academic writing task by improving writing knowledge gains, fostering self-regulated learning (SRL) skills and building different human-AI dialogue characteristics. Grounded in Zimmerman's SRL framework, the LAD provided real-time feedback on learners' goal-setting, writing processes and reflection, while monitoring the quality of learner-AI interactions. A quasi-experiment was conducted involving 52 postgraduate students divided into an experimental group (EG) using the LAD to a control group (CG) without it in a human-AI collaborative writing task. Pre- and post- knowledge tests, questionnaires measuring SRL and cognitive load, and students' dialogue data with GenAI were collected and analyzed. Results showed that the EG achieved significantly higher writing knowledge gains and improved SRL skills, particularly in self-efficacy and cognitive strategies. However, the EG also reported increased test anxiety and cognitive load, possibly due to heightened metacognitive awareness. Epistemic Network Analysis revealed that the EG engaged in more reflective, evaluative interactions with GenAI, while the CG focused on more transactional and information-seeking exchanges. These findings contribute to the growing body of literature on the educational use of GenAI and highlight the importance of designing interventions that complement GenAI tools, ensuring that technology enhances rather than undermines the learning process.

# 1. Introduction

The rapid emergence of Generative AI (GenAI) has sparked considerable interest among researchers and educators, primarily due to its transformative potential in reshaping learning processes. GenAI's advanced capabilities—such as generating human-like text, solving complex problems, and delivering personalized feedback—have introduced innovative opportunities to enhance educational practices (Chang et al., 2024; Jia et al., 2024; Urban et al., 2024). Nevertheless, recent research has unveiled a more complex and concerning dimension of GenAI's role in education. A central issue identified in the literature is the risk of over-reliance on GenAI, which may impede the development of students' independent problem-solving skills and critical thinking abilities (Fan et al., 2024; Gerlich, 2025). For instance, a current study (Bastani et al., 2024) conducted at the University of Pennsylvania involving nearly 1,000 students in grades 9–11 over one semester revealed a paradoxical effect of GenAI use. While students who utilized ChatGPT for review guidance demonstrated short-term grade improvements, their performance in subsequent closed-book tests declined markedly. Lehmann et al. (2025) found that there was no effect of GenAI on learners' overall learning outcomes, and GenAI could widen the gap between students with low and high prior knowledge. Those findings suggest that GenAI may foster superficial learning and cognitive offloading (Gerlich, 2025), ultimately undermining long-term academic retention and mastery.

In this situation, some researchers started to explore effective human-AI collaboration in learning context. Scholars have introduced the notion of hybrid intelligence, emphasizing the complementary strengths of human and artificial intelligence that foster mutual learning and ongoing enhancement (Molenaar, 2022a, 2022b). According to Dellermann et al. (2019), hybrid intelligence as the integration of human and AI capabilities to achieve complex objectives, producing better results than individual efforts while enabling continuous mutual learning. In an optimal setup, humans and AI share agency, each leveraging their distinct competencies to achieve the best possible results (Cukurova, 2024; Järvelä et al., 2023). As Molenaar's (2022a) conceptual framework of Hybrid Human-AI Regulation (HHAIR) suggests, learners' self-regulation skills play a critical role in facilitating effective human-AI collaboration. However, current discussions on human-AI collaboration in learning remain largely conceptual, particularly in the context of writing learning. There is a notable lack of empirical research exploring how to promote such collaborative relationships and enable learners to fully benefit from interactions with advanced GenAI systems.

To address these challenges, learning analytics dashboards (LADs) have emerged as a promising tool to facilitate effective human-AI collaboration. LADs are designed to provide supports by visualizing learning processes and progress, thereby empowering learners to monitor, reflect on, and regulate their learning behaviors (Bodily, Kay, et al., 2018; Jivet et al., 2018), which could serve as a mediating tool to enhance learners' metacognitive awareness and facilitate more intentional interactions with AI systems (Jadhav et al., 2024). However, as previous research pointed out, existing LADs primarily focus on the teacher's perspective and are mostly used for data visualization rather than being theory-driven, which could not turn the learning analytics into actional insights for learners (Matcha et al., 2020; Paulsen & Lindsay, 2024). Moreover, currently there has raising research discussed the gap between the design of conventional LAD and future classrooms integrated with ChatGPT. For example, it is unclear how a LAD can gain an in-depth understanding of learners' behavior, prevent learners' over-reliance and misuse, and facilitate effective use of ChatGPT (Kim et al., 2024a, 2024b).

Therefore, this study aims to investigate the role of a theory-driven learning analytics dashboard in optimizing human-AI collaboration for learning academic writing. Specifically, the study examines whether a LAD grounded in Zimmerman's (2001) self-regulated learning (SRL) framework can (1) enhance learners' writing knowledge rather than merely improving task performance; (2) foster SRL skills while managing the consequent cognitive load and (3) influence dialogue interaction between learners and GenAI. The findings aim to provide insights for enhancing human-AI collaboration in writing learning that exploring how can GenAI augments learning outcomes rather than undermines it.

## 2. Background

**2.1 Concerns over GenAI assistance in educational context**

Previous studies have demonstrated that GenAI can enhance learners' performance across various educational tasks, offering scalable and adaptive support that complements traditional teaching methods. For instance, Urban et al. (2024) found that ChatGPT's assistance significantly improved the quality, refinement, and originality of students' solutions in problem-solving tasks. Similarly, Meyer et al. (2024) reported that students achieved higher writing performance when receiving feedback generated by GenAI on essay writing tasks compared with those receiving no feedback. This highlights the potential of GenAI to provide timely, detailed, and actionable feedback. These findings l with the growing body of literature suggesting that GenAI can effectively implement feedback mechanisms in educational settings, thereby supporting students in improving their learning performance (Chang et al., 2024; Vargas-Murillo et al., 2023; Zhang & Tur, 2024).

However, the integration of GenAI in education is not without its challenges. Despite its potential benefits, some studies cautioned that these improvements may not translate into meaningful gains in learners' underlying knowledge or understanding. For example, Fan et al. (2024) observed that although ChatGPT significantly improved students' academic writing performance, it did not lead to substantial gains in knowledge acquisition. This suggests that GenAI's effectiveness may be limited to surface-level improvements, where it can optimize content and enhance performance quickly, particularly in tasks with clear, structured requirements and scoring criteria (Steiss et al., 2024). In such cases, GenAI acts as a powerful tool for task completion but may not foster deeper learning or critical thinking (Gerlich, 2025). Conversely, tasks that require higher-order thinking skills, such as independent problem-solving, effective help-seeking, and knowledge transfer, may not benefit as much from GenAI assistance (e.g., Author et al., 2024; Cooper, 2023; Fan et al., 2024). Furthermore, Lo et al. (2024) highlighted behavioral concerns, noting that while many students actively use ChatGPT in learning activities, misuse of the tool—often stemming from insufficient digital literacy and understanding—can lead to issues such as plagiarism and academic dishonesty. These findings underscore the concern that GenAI is helping learners improve task performance rather than their actual abilities, potentially fostering a superficial understanding of knowledge. As Yan et al. (2024) have emphasized, current studies should prioritize understanding how to enhance learners' actual knowledge gains, bringing pedagogical interventions into GenAI learning environment, and ensuring that GenAI serves as a powerful ally in the pursuit of knowledge and innovation rather than a crutch that undermines intellectual abilities.

## 2.2 Human-AI collaboration in writing and self-regulated learning

Writing, a cognitively demanding task, involves planning, translating, and reviewing stages, requiring skills like idea expression, logical structuring, and perspective-shifting (Graham, 2018; Hayes & Flower, 1986). Traditional AI-assisted writing has primarily focused on improving mechanical aspects like grammatical accuracy (McCarthy et al., 2022), spelling correctness (Parra G & Calero S, 2019), and coherence (Chen & Cui, 2022) through natural language processing (NLP) tools that detect and correct errors instantly, aligning with the broader efficacy of automated writing evaluation (AWE) systems in enhancing linguistic accuracy (Author, et al., 2024). However, this model of traditional AI collaboration focuses on error reduction rather than fostering deeper cognitive engagement in writing. The advent of GenAI has reshaped the landscape of human-AI collaboration in writing, particularly in higher education contexts where students increasingly leverage its text-generation capabilities to enhance their writing processes (Barrett & Pack, 2023; Li et al., 2024). Unlike earlier AI assistance, GenAI facilitates a more dynamic and interactive relationship between learners and AI, enabling tasks such as generating initial drafts, summarizing literature, and even proposing creative content (Barrett & Pack, 2023; Li et al., 2024). Recent studies reveal that active engagement with AI-generated text—where learners critically evaluate, adapt, and build upon AI outputs—can foster creativity, coherence, and lexical precision (Doshi & Hauser, 2024). Conversely, passive reliance on AI increases dependency and undermines the development of essential writing skills (Fan et al., 2024). This distinction aligns with Bereiter and Scardamalia's (2013) framework of knowledge building, which differentiates between knowledge telling—a linear process of transcribing information without deep engagement—and knowledge transformation—a more critical process that involves restructuring arguments, synthesizing viewpoints, and creating cohesive narratives. From this perspective, the effectiveness of human-AI collaborative writing hinges on the dynamic interplay between human and AI capabilities. Tasks can be offloaded to AI when suitable (e.g., generating initial drafts or summarizing literature) but should be reassumed by humans when complexity demands critical engagement (Cheng et al., 2024; Wan et al., 2024).

As previous research has pointed out, at the heart of this dynamic interplay lies the concept of self-regulated learning (Cukurova, 2024; Molenaar, 2022a). SRL empowers learners to take control of their learning processes by setting goals, planning strategies, monitoring progress, and reflecting on outcomes (Zimmerman, 2001). In the context of human-AI collaborative writing, SRL skills ensure that learners actively engage with AI-generated content, critically evaluate its relevance, and integrate it meaningfully into their knowledge base (Darvishi et al., 2023; Fan et al., 2024). For example, while AI can assist in summarizing literature or generating drafts, it is the learner's ability to self-monitor and regulate their learning that ensures the final output is coherent, accurate, and reflective of deep understanding (Nguyen et al., 2024). However, empirical research has primarily examined SRL in traditional AI-assisted writing context. For example, Sari and Han (2024) found that there was no statistically significant difference between the AWE feedback and teacher feedback in promoting the use of self-regulated writing strategies by the students, while Yabukoshi and Mizumoto (2024) proposed an SRL strategy-based instruction for AI-assisted writing and showed statistically significant improvements in students' SRL strategy use. But currently little research on the context of collaborative writing with GenAI ones. This gap is significant because current human-AI collaborative writing inherently involves a nuanced interplay between human and GenAI contributions, demanding that learners develop hybrid regulatory strategies to align AI's capabilities with their cognitive goals (Järvelä et al., 2023). Without a deliberate focus on fostering SRL in GenAI collaboration contexts, the increasing delegation of tasks to AI risks can undermine learners' abilities in critical thinking and cognitive engagement (Gerlich, 2025).

**2.3 Effect of learning analytics dashboard in supporting learning**

Learning analytics dashboards have emerged as a promising tool to foster metacognition and SRL by visualizing real-time data to enhance students' awareness of their learning processes (Bodily, Ikahihifo, et al., 2018; Chen et al., 2021). Historically, LADs have been designed to support discrete SRL components, such as goal-setting and planning (e.g., assignment schedules in Bodily et al., 2018), progress monitoring (e.g., resource access views in Molenaar et al., 2019), and reflection (e.g., grade distributions and peer comparisons in Kim et al., 2016). While these studies suggest LADs can positively influence specific SRL skills, their focus on isolated modules rather than a holistic, theory-driven SRL framework raises significant concerns. Past reviews underscore that most LADs lack grounding in established learning theories (Matcha et al., 2020), prioritizing technical data visualization over pedagogical coherence. This fragmented approach limits their capacity to scaffold the full spectrum of SRL processes—planning, monitoring, and reflection—potentially reducing their effectiveness in fostering meaningful learning behaviors.

The absence of a comprehensive SRL theoretical foundation in LAD design not only constrains their pedagogical impact but also risks increasing cognitive load, particularly in complex tasks like writing. Writing demands navigating multiple cognitive processes—planning, translating, and reviewing—each reliant on working memory capacity and sensitive to task complexity (Graham, 2018). Sweller's (1994) cognitive load theory posits that extraneous load, such as that induced by poorly integrated or overly technical LADs, can overwhelm learners, hindering these processes. For instance, Ferguson and Clow (2017) and Guzmán-Valenzuela et al. (2021) caution that certain LAD designs, by presenting data without actionable insights, may exacerbate cognitive strain rather than alleviate it. This issue is compounded by learners' varying data literacy levels (Gebre, 2022) and the potential for LADs to deliver feedback that is difficult to translate into practical strategies (Sedrakyan et al., 2020). Recent studies further suggest that such designs can elevate cognitive load, undermining SRL efforts (Cheng et al., 2024), especially when students lack the skills to interpret complex visualizations effectively.

In the context of GenAI-assisted learning environments, these challenges become even more pronounced, yet evidence remains inconclusive (Paulsen & Lindsay, 2024; Valle et al., 2021). While LADs hold potential to support SRL in human-AI collaborative settings—such as writing with GenAI tools—little is known about how theory-driven LADs can mitigate cognitive load and enhance SRL dimensions like planning, monitoring, and reflection. The lack of theoretically grounded LADs tailored to these emerging scenarios limits their integration into educational practice, as they fail to address the dynamic interplay between AI assistance, learner agency, and cognitive demands (Sedrakyan et al., 2020). Consequently, further investigation is needed to design LADs that not only support isolated SRL modules but also align with comprehensive SRL frameworks, reducing cognitive load and optimizing learner interactions in GenAI-augmented contexts.

**2.4 Research Questions**

Building on previous insights of research, this study seeks to explore the role of learning analytics dashboard in human-AI collaborative writing contexts. Specifically, it aims to address the following research questions:

RQ1: Can the theory-driven learning analytics dashboard enhance learners' writing knowledge gains in the human-AI collaborative writing context?

RQ2: How does the theory-driven learning analytics dashboard influence learners' SRL and cognitive load in the human-AI collaborative writing context?

RQ3: What is the impact of theory-driven learning analytics dashboard on learners' dialogue interaction with GenAI in the human-AI collaborative writing context?

# 3. Methodology

## 3.1 Participants

This study employed a quasi-experimental design to investigate the effect of a learning analytics dashboard on the human-AI collaborative writing task. An a priori power analysis was conducted using G*Power to determine a suitable sample size for detecting a large effect (Cohen's d = 0.8) with 75% power at a significance level of $\alpha = .05$ (two-tailed), assuming an independent samples t-test for group comparison. The analysis indicated that a minimum total sample size of 40 participants (20 per group) would be required. The study was conducted at a university in China, where 52 students in two classes enrolled in the "Academic Writing" course with the same lecturer were recruited. These participants had a mean age of 23.1 years, with the sample including 20 females and 32 males. All students were majoring in STEM disciplines (such as computer science, electronics, and communication engineering) and all reported have prior experience in using GenAI. One class was assigned as the experimental group (EG, N=27, 11 females and 16 males), while the other served as the control group (CG, N=25, 9 females and 16 males). All students were informed of it was an educational experiment and signed the informed consent voluntarily.

**3.2 Research Design and Procedure**

The study focused on an academic abstract writing task in which students were required to read the provided academic articles and compose corresponding abstracts. Our study was conducted in the middle of the term. All the participants had taken academic abstract writing classes prior to our experiment task to ensure they had adequate foundational knowledge.

The experiment followed a pre-test/post-test design with a total duration of 130 minutes (see Figure 1 for the completed procedure) in the "Academic Writing" course. In the initial 15-minute pre-test phase, both groups completed a knowledge test to assess their understanding of academic writing and a questionnaire to evaluate their SRL skills and cognitive load. During the 100-minute intervention phase, both groups engaged in an academic abstract writing task that involved reading and writing while collaborating with ChatAI (See Figure 2). The backend large language model of ChatAI was ChatGLM-4 model, an open-source large language model with performance comparable to GPT-4o (GLM Team et al., 2024). The reading material was an academic paper examining the effects of background music on preschoolers' cognitive task performance (Koolidge & Holmes, 2018). We removed the original abstract and conclusion sections, requiring students to read the main body of the paper and write a new abstract in our developed writing interface. The writing interface included a text-entry area and a toolbar with basic formatting functions, such as font selection, text alignment, and list creation. Throughout the reading and writing process, both groups had access to a ChatAI module where they could interact with GenAI for reading comprehension and writing assistance. To foster meaningful human-AI collaboration while preventing over-reliance on AI-generated content, we implemented two key constraints for both groups: First, the ChatAI system was configured to decline responses to queries exceeding 30 words, preventing students from requesting whole-paper translation or whole-paper automatic summarization. The threshold of 30 words was determined by previous pilot study that most of questions (near 80%) of students in this task were less than 30 words. Second, the writing interface disabled the "paste" function, preventing students from directly copy the GenAI-generated content to paste in their writing output. Throughout the experiment, participants in both groups could access previous course materials, including PowerPoint slides detailing academic abstract writing guidelines. The experiment concluded with a 15-minute post-test phase, which mirrored the pre-test structure with a knowledge test and questionnaire test to evaluate writing knowledge, SRL and cognitive load.

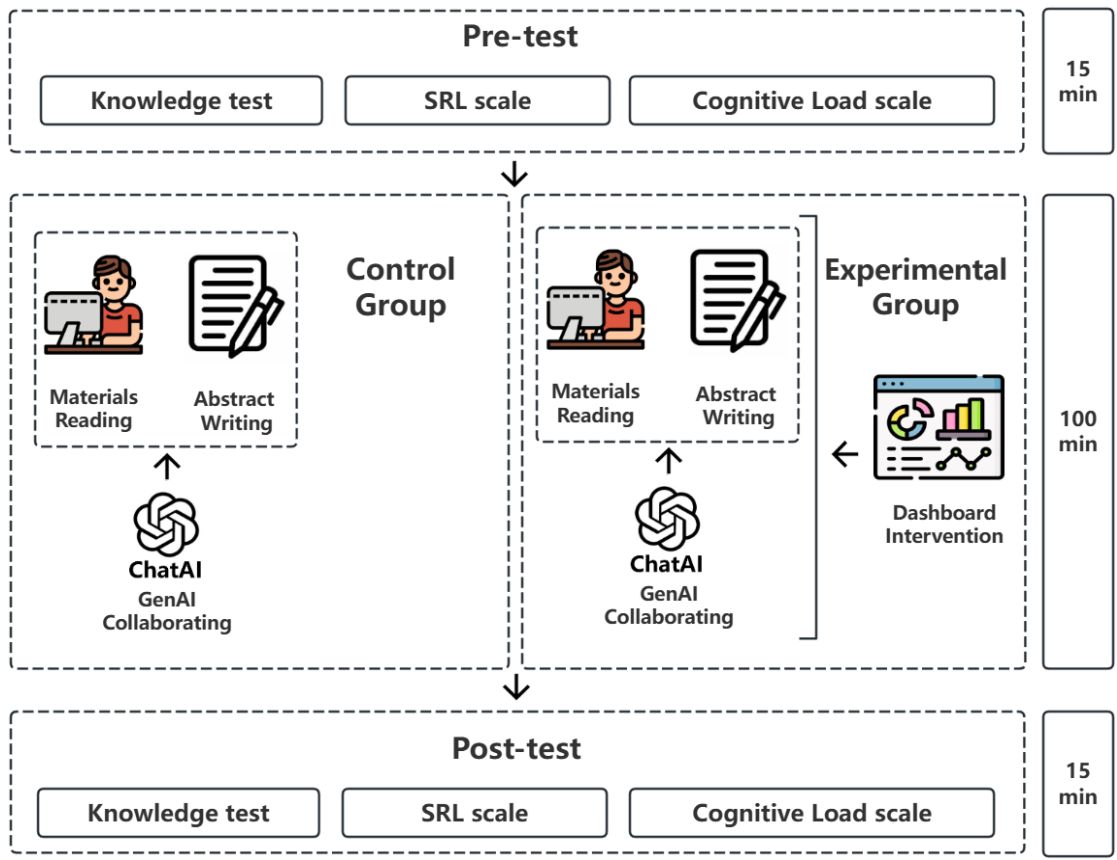

Figure 1 Experiment Procedure

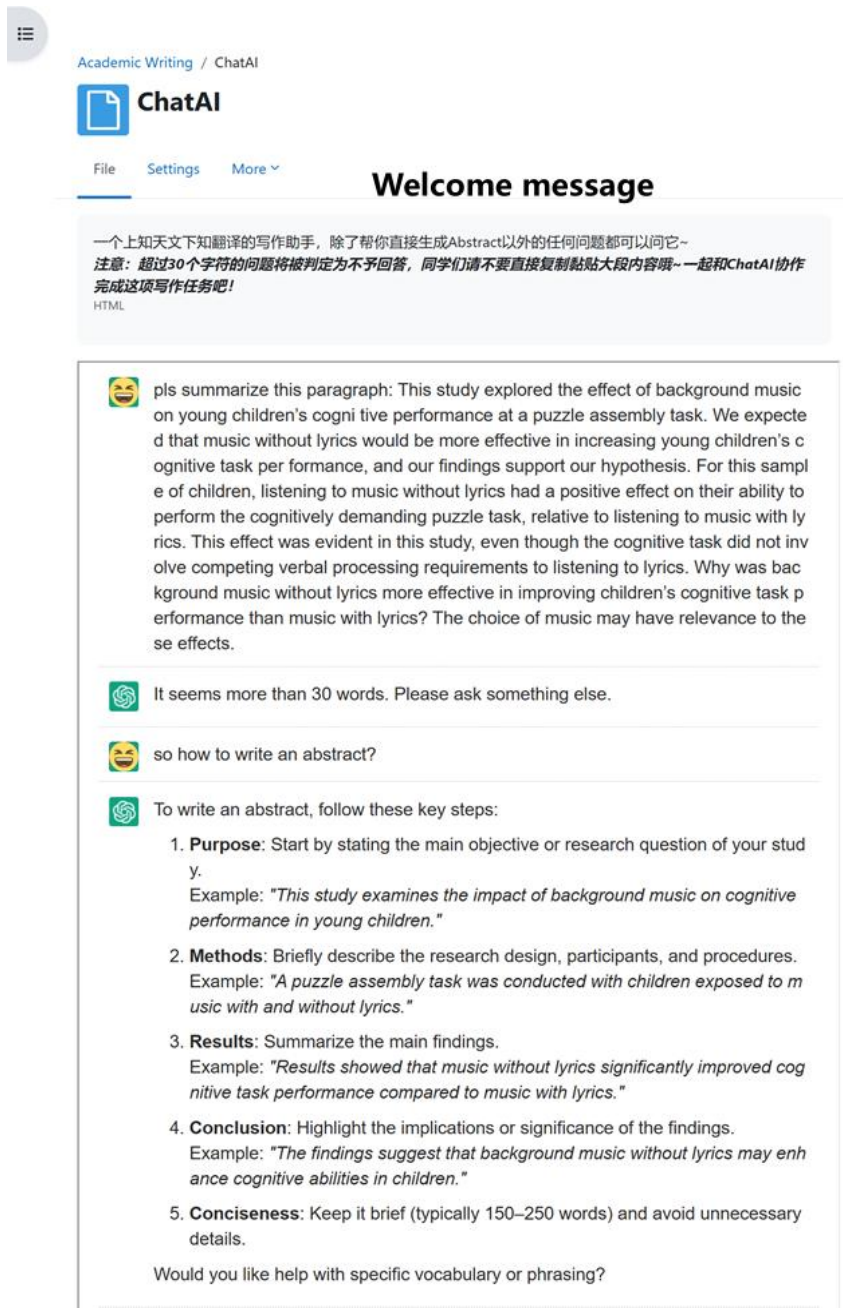

Figure 2 "ChatAI" GenAI chatbot interface used in our experiment

## 3.3 Learning analytics dashboard design in human-AI collaborative writing context

In this study, we designed a learning analytics dashboard based on the three-phase Self-Regulated Learning model (Zimmerman, 2001) (See Figure 3). Specifically, the dashboard's primary features aligned with the SRL phases: First, the Writing Preparation Module aligns directly with the forethought phase of Zimmerman's (2001) SRL model, which emphasizes processes that precede action, such as goal setting and strategic planning. This module operationalized this phase by presenting a pentagon radar chart for goal setting across five critical dimensions of academic writing: expected time, content understanding, structure completeness, expression accuracy, and logical coherence. These dimensions were chosen based on criteria of the IELTS writing band (Read, 2022) . Learners were required to set goal scores for their abstract writing output (e.g., the expected structure completeness score of their writing). Detailed explanations for each dimension were provided to help learners set their writing goals (See Appendix 1). The theoretical significance of this module lies in its attempt to translate a core component of SRL theory into a practical step for learners. By encouraging students to clearly define their goals and standards before they start the writing task, the Writing Preparation Module aimed to help them become more aware of their own thinking processes (metacognitive awareness) and better able to guide their own learning right from the beginning (Azevedo, 2009). This, in turn, helps build a strong and organized foundation so that students can more effectively manage their own learning during the later stages when they are writing (performance phase) and when they think back on their writing afterwards (self-reflection phase).

Second, the Writing Process Monitoring Module directly targeted the performance phase of Zimmerman's (2001) SRL model. This phase is critical as it involves the active execution of tasks alongside self-monitoring of progress and the strategic control of learning processes. The module supported this by displaying a real-time analysis of academic abstract component completeness, including research background, question, method, results, and conclusion. This analysis, presented on a 5-point scale (from "not begun" to "fully completed"), allowed learners to continuously observe and track their progress in constructing a well-structured abstract while actively engaged in the writing process. The real-time analysis and scoring were conducted by the GenAI model ChatGLM-4. The choice of GenAI for this function was informed by evidence demonstrating its accurate assessment capabilities in writing tasks, comparable to human evaluators (e.g., Seßler et al., 2025; Steiss et al., 2024; Yoshida, 2024), and GenAI showed its high usability serve as a learning analytics tool (Yan, Martinez-Maldonado, et al., 2024). The prompt for scoring the completeness of each dimension can be found in the Appendix 2. By giving students instant and continuous feedback on how complete their writing structure is, this module puts the important self-monitoring part of the performance phase into action (Matcha et al., 2020). This ongoing feedback helps learners see any gaps between their current writing and what a full abstract requires. As a result, it encourages them to make timely changes to their approach, such as focusing more effort on unfinished sections or improving what they have already written.

Third, the Writing Reflection Module directly targeted the self-reflection phase of Zimmerman's (2001) SRL model. This phase is critical as it involves learners evaluating their performance, making judgments about the causes of their outcomes (causal attributions), and reacting in ways (self-reactions) that influence their motivation and future learning strategies. The module supported this by providing a comparative radar chart that overlaid learners' initial goals (set in the Writing Preparation Module) with their actual writing output scores on the corresponding dimensions (also assessed by the ChatGLM-4 model). The assessment prompt for the model can also be found in the Appendix 3. By presenting this direct comparison of goals and outcomes, the module operationalized the self-judgment component of the self-reflection phase, specifically prompting self-evaluation against self-set standards. The visual representation of the gap between their desired and actual performance was designed to stimulate causal attributions, encouraging learners to reflect on why they succeeded or failed in certain areas (e.g., "I scored lower on 'logical coherence' because I didn't plan my paragraph transitions well"). These reflections are crucial for informing their self-reactions (e.g., satisfaction with progress, or recognition of a need for strategic change) and guiding adaptive inferences for subsequent writing tasks, such as adjusting future goals or planning different strategies.

Fourth, to help learners better collaborate with ChatAI in their academic writing, our dashboard included an AI Dialogue Quality Module, which also supported the self-reflection phase by focusing on the metacognitive aspects of human-AI interaction. Zimmerman's self-reflection phase also encompasses evaluating the effectiveness of the strategies employed during the performance phase (Zimmerman, 2001). This module analyzed students' latest five questions posed to ChatAI, assessing dimensions such as task focus, question quality, academic standards, independent thinking, and questioning strategy on a 0-100 scale (also assessed by the ChatGLM-4 model). The assessment prompt for the model can also be found in the Appendix 4. By providing feedback on the quality of their interactional strategies with the GenAI, this module prompted learners to engage in self-judgment regarding the efficacy of their help-seeking behaviors and collaborative approaches. For instance, a low score on 'independent thinking' might lead a student to reflect on their over-reliance on AI for ideas. This process encourages causal attributions related to their interaction patterns (e.g.,

"My questions were too vague, so the AI's help wasn't optimal") and fosters adaptive inferences for future collaborations, such as formulating more precise prompts or critically evaluating GenAI suggestions more thoroughly, thereby promoting more effective co-regulation and deeper engagement.

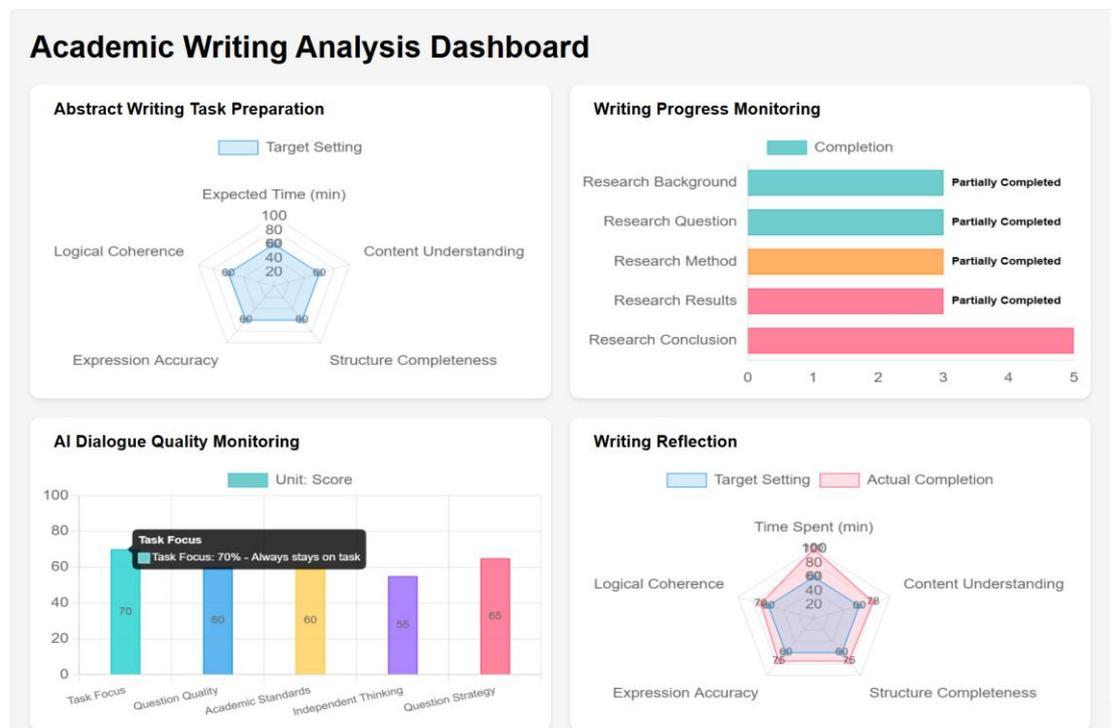

Figure 3 Learning analytics dashboard based on three-phase SRL model

As recommended by Khosravi et al. (2022), explainable AI has become increasingly important for AIED systems currently. Understanding the decisions made by AI helps students better trust AI to guide their learning behaviors. And previous research showed providing actionable explanations in AIED systems can promote learning outcomes (Winne, 2021). In our study, to help students better understand the AI analyzed results of each module, we designed explanatory pages for each module in our dashboard (See Figure 4). These pages would pop up when students click on the corresponding modules. The explanatory pages explain what the scores represent for each dimension, and display the reasons why the ChatGLM-4 model assigned certain scores. The text for these explanations is generated by the ChatGLM-4 model simultaneously as it assesses the scores of different dimensions. Before the experiment began, all students read an introductory document about these pages and were given time to familiarize themselves with the interface the day prior. One of the authors resolved all queries from students to ensure they had a basic familiarity with the feature during the experiment. Finally, to minimize potential hallucinations from the GenAI system, we set its "temperature," "frequency penalties," and "presence penalties" parameters to zero. This ensures that the large language model chooses the highest-probability tokens, resulting in consistent assessment results and scores.

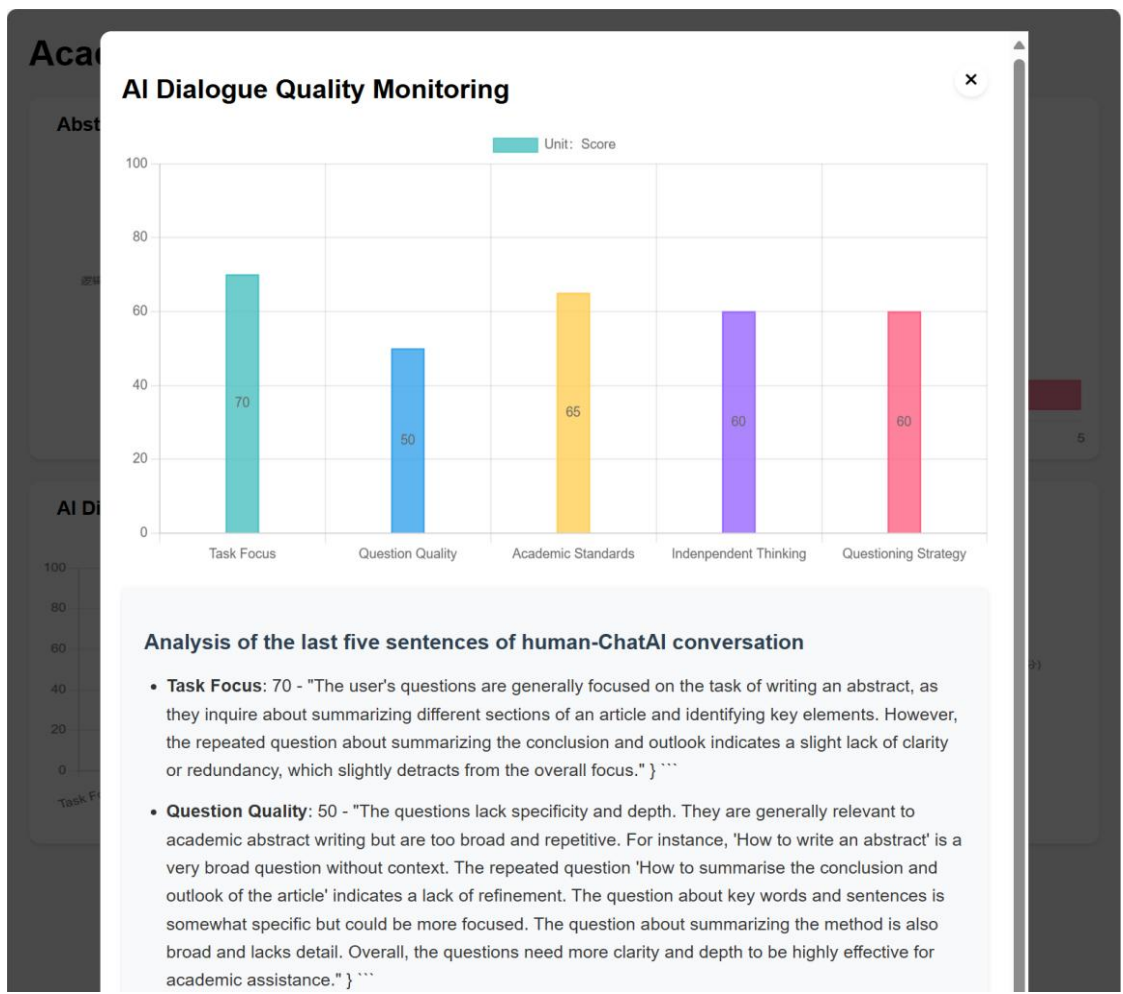

Figure 4 Example of explanatory page for AI Dialogue Quality Monitoring module

## 3.4 Data Collection and Analysis

For RQ1, to examine the effects of the learning analytics dashboard intervention on students' academic writing knowledge, we collected data at two time points (pre-test and post-test). The knowledge test, designed by an English lecturer with 10 years of teaching experience in academic writing, assessed students' understanding of an abstract's structural components, academic writing conventions, formal language usage, and other academic writing knowledge in 20 single-choice questions. The full test is provided in Appendix 6. This test served as both a pre-test to assess initial group equivalence on academic writing knowledge and a post-test to measure learning gains after the intervention. To analyze the impact on academic writing knowledge, a 2 (group: experimental vs. control) × 2 (time: pre-test vs. post-test) mixed-design Analysis of Variance (ANOVA) was first conducted. This analysis was chosen as the primary method to assess changes in knowledge over time and to determine if these changes differed between the experimental and control groups, specifically by examining the time × group interaction effect.

For RQ2, to comprehensively evaluate the impact of the dashboard on students' SRL and cognitive load, our study employed several questionnaires. First, SRL ability was measured using an adapted version of the Motivated Strategies for Learning Questionnaire (MSLQ; Pintrich et al., 1991), which is one of the most widely used and validated instruments for assessing self-regulated learning (Tock & Moxley, 2017). Following a previous study's adaptation (Liao et al., 2024), we focused on four key subscales of MSLQ: *self-efficacy* (8 items, α=0.839), *intrinsic motivation* (4 items, α=0.873), *test anxiety* (5 items, α=0.737), and *cognitive strategies* (12 items, α=0.896). All items were rated on a 7-point Likert scale ranging from 1 (not at all true of me) to 7 (very true of me). The self-efficacy subscale measures students' beliefs about their capability to perform the English academic writing task; the intrinsic motivation subscale assesses students' inherent interest and engagement in the learning process; the test anxiety subscale evaluates students' emotional and cognitive concerns about task performance; and the cognitive strategies subscale examines students' use of learning strategies in the writing process. Second, cognitive load was measured using an adapted version of the Cognitive Load Scale (Paas, 1992; Wang et al., 2023). To distinguish between expectations and actual experience, the items were framed differently at each time point. At pre-test, students rated their agreement with prospective statements to assess their anticipated cognitive load (e.g., "I expect I will need to invest a lot of mental effort in this abstract writing activity."). At post-test, they rated retrospective statements to assess their perceived cognitive load experienced during the task (e.g., "I invested a lot of mental effort in this abstract writing activity."). The scale consists of two dimensions, mental load and mental effort, with four seven-point Likert rating items in each (α=0.727 for pre-test mental load; α=0.857 for post-test mental load; α=0.790 for pre-test mental effort; α=0.731 for post-test mental effort; details and the full list of items can be found in Appendix 7).

For RQ3, we collected dialogue data between learners and ChatAI from our log data to further investigate the dialogue characteristics of different groups. These dialogues were analyzed using a coding framework adapted from Cheng et al., (2025). In their study, learners' questions to GenAI in an essay reading and writing task were categorized into shallow and deep depth, with fourteen distinct types identified. We modified their framework by refining the detailed definitions and providing relevant examples tailored to our specific task (See Table 1). Two coders independently coded 50% of the data using the framework, achieving high inter-coder reliability (Cohen's Kappa = 0.77). Conflicting codes were resolved through discussion to reach consensus. Subsequently, one coder completed the remaining coding based on the framework.

Table 1 Adapted coding framework of human-AI question-asking types from Cheng et al., (2025)

| Question Type | Depth | Definition | Examples |
|---|---|---|---|
| Verification | Shallow | Questions to confirm the truth or occurrence of a fact or event | You mean the first sentence still needs to expand? |
| Disjunctive | Shallow | Questions to determine which among a set of options is the case | Should I first do x, or do y? |
| Concept Completion | Shallow | Questions to identify or complete a missing element, usually a referent of a noun argument slot | What is the correct spelling of Mozart effect? |
| Example | Shallow | Questions to identify an instance or label that exemplifies a category | Can you give examples on how to write a conclusion in abstract? |
| Feature Specification | Shallow | Questions to understand the qualitative attributes of an entity | Can you explain what is classical music? |
| Definition | Shallow | Questions to clarify the meaning of a term or concept | What is the definition of Mozart effect? |
| Comparison | Shallow | Questions to explore similarities and differences between two or more entities | What's the difference between academic result and conclusion? |
| Casual Consequence | Deep | Questions to understand the effects of an event or state | What are the potential consequences of not providing background in abstract? |
| Instrumental | Deep | Questions to identify the means or methods to accomplish a goal | How to use past tense in abstract writing? |
| Enablement | Deep | Questions to understand the resources or conditions that allow an action to be performed | Can you give me some ideas of specific format requirements, such as word count? |
| Judgmental | Deep | Questions to evaluate an idea or to seek advice | How would you rate the quality of an abstract? |

| Assertion | Not Specified | Question that indicates a lack of knowledge or understanding of an idea | uh, i don't know what revision is needed to improve my abstract haha |
|---|---|---|---|
| Indirect Request | Not Specified | Questions asked in a polite and indirect form when the speaker wants the listener to perform a specific action | Could you please give me the feedback of my abstract? |
| Direct Request | Not Specified | Questions asked in a commanding or direct form when the speaker wants the listener to perform a specific action | Give me the feedback of my abstract. |

We employed Epistemic Network Analysis (ENA) to explore structural differences in learners' question-asking types when interacting with ChatAI between groups. ENA is a method that visualizes the relationships and connections between different concepts and behaviors in a network, making it particularly useful for identifying patterns in cognitive processes (Shaffer et al., 2016). After coding the data using the framework described above, the results were imported into the ENA web tool (http://www.epistemicnetwork.org/), which calculated how each message was connected to previous ones, in what is called stanzas. Specifically, ENA uses a moving window to divide a content-related multi-line discourse message into a stanza. In this study, as determined by the continuity of the participants' online messages in this discourse dataset, relevant prior work (e.g., Sun et al., 2023), and by our own finding that the most meaningful connections in the discourse occurred within the space of four messages, the stanza-window size was set at four.

## 4. Result

**4.1 Writing Knowledge Tests**

Descriptive statistics for the writing knowledge test scores are presented in Table 2. A 2 (group: Experimental, Control) × 2 (time: Pre-test, Post-test) mixed-design ANOVA was conducted to examine the effects of the learning analytics dashboard on students' academic writing knowledge (See Table 3). The ANOVA revealed no significant main effect of time, $F(1, 50) = 0.095$, $p = .759$, but a significant main effect of group, $F(1, 50) = 4.477$, $p = .041$. More importantly, a significant time × group interaction effect was found, $F(1, 50) = 5.616$, $p = .023$. This significant interaction indicates that the change in academic writing knowledge scores from pre-test to post-test differed significantly between the experimental and control groups. To decompose this interaction, post-hoc pairwise t-tests with Bonferroni correction were performed. Results showed that the experimental group (EG) demonstrated a significant improvement in academic writing knowledge from pre-test (M = 81.36, SD = 11.72) to post-test (M = 90.00, SD = 6.90), t = -3.47, adjust $p$ = 0.002. In contrast, the CG showed no significant change from pre-test (M = 80.14, SD = 10.23) to post-test (M = 82.70, SD = 14.4), t = 1.01, adjust $p$ = 0.33. Furthermore, at post-test, the EG scored significantly higher than the control group (t = 2.69, adjust $p$ = .011). Notably, there were no significant differences between the groups at pre-test, t = 0.469, adjust $p$ = .642, confirming baseline equivalence. These findings indicate that students who used the learning analytics dashboard intervention (EG) achieved greater improvement in academic writing knowledge than those without the dashboard support (CG).

Table 2 Descriptive Statistics for Academic Writing Knowledge Test Scores (Mean ± SD)

| Group | N | Pre-test Scores (M±SD) | Post-test Scores (M±SD) |
|---|---|---|---|
| CG | 27 | 80.14±10.23 | 82.70±14.4 |
| EG | 25 | 81.36±11.72 | 90.00±6.90 |

Table 3 mixed-design ANOVA results

| Effect | df_num | df_den | F | p | $\eta^2$ |
|---|---|---|---|---|---|
| Group | 1 | 50 | 4.477* | 0.041 | .051 |
| Time | 1 | 50 | 0.095 | 0.759 | .001 |
| Time × Group | 1 | 50 | 5.616* | 0.023 | .078 |

Note: * p<0.05.

### 4.2 SRL skills

Quantitative analysis of SRL dimensions revealed differential effects of the dashboard intervention across subscales (Table 3). No significant baseline differences were found between groups in pre-test SRL measures. Due to non-normal distributions identified through Mann-Whitney U tests ($p < 0.05$ for all subscales), between-group comparisons of post-test scores were analyzed using Mann-Whitney U tests with Cliff's delta effect size calculations.

Results showed that EG demonstrated statistically significant enhancements in three critical SRL components. For self-efficacy, EG participants (M = 6.07, SD = 0.50) outperformed their CG counterparts (M = 5.45, SD = 0.72) with moderate effect size (U = 74, Z = -2.737, $p$ = 0.005, ES =

-0.537), indicating substantially stronger confidence in their academic writing capabilities. Similarly, the EG exhibited higher cognitive strategy utilization (M = 5.60, SD = 0.67) than the CG (M = 4.92, SD = 0.57; U = 72.5, Z = -2.785, $p$ = 0.005, ES = -0.546), suggesting more effective deployment of learning strategies during the writing process.

Interestingly, test anxiety scores showed a counterintuitive pattern, with the EG reporting higher anxiety levels (M = 5.02, SD = 1.08) than the CG (M = 4.04, SD = 1.33) at a statistically significant level (U = 93, Z = -2.132, $p$ = 0.033, ES = -0.418). No significant between-group differences emerged in intrinsic motivation (U = 142, Z = -0.573, $p$ = 0.569), suggesting that the intervention primarily affected regulatory processes rather than fundamental motivation levels.

Table 3 Comparison of students' SRL skills between CG and EG

| Dimension | CG Group M(SD) | EG Group M(SD) | U | Z | $p$ | Effect Size |
|---|---|---|---|---|---|---|
| Self-Efficacy | 5.45(0.72) | 6.07(0.50) | 74 | -2.737 | 0.005** | -0.537 |
| Intrinsic Motivation | 5.54(0.91) | 5.63(1.15) | 142 | -0.573 | 0.569 | -0.112 |
| Test Anxiety | 4.04(1.33) | 5.02(1.08) | 93 | -2.132 | 0.033* | -0.418 |
| Cognitive Strategy | 4.92(0.57) | 5.60(0.67) | 72.5 | -2.785 | 0.005** | -0.546 |

Note：*** $p$<0.001; ** $p$<0.01; * $p$<0.05

## 4.3 Cognitive Load

Table 4 Comparison of students' cognitive load between CG and EG

| Dimension | CG Group M(SD) | EG Group M(SD) | U | Z | $p$ | Effect Size (ES) |
|---|---|---|---|---|---|---|
| Mental Load | 5.45(0.72) | 6.07(0.50) | 78 | -2.611 | 0.008** | -0.512 |
| Mental Efforts | 5.33(0.68) | 5.45(0.76) | 144.5 | -0.493 | 0.598 | -0.096 |

Note：*** $p$<0.001; ** $p$<0.01; * $p$<0.05

The Mann-Whitney tests were also conducted for the analysis of cognitive load due to the non-normal distributions (Table 4). For mental load, the EG reported significantly higher scores (M = 6.07, SD = 0.50) compared with the CG (M = 5.45, SD = 0.72), with a statistically significant difference (U = 78, Z = -2.611, *p* = 0.008, ES = -0.512). This moderate effect size suggests that the dashboard intervention might introduce additional cognitive demands, possibly from the real-time monitoring and self-regulation processes facilitated by the SRL-aligned dashboard components (e.g., goal-setting radar charts, dynamic feedback on abstract components, and writing reflections). In contrast, no significant difference emerged in mental effort (EG: M = 5.45, SD = 0.76; CG: M = 5.33, SD = 0.68; U = 144.5, Z = -0.493, *p* = 0.598, ES = -0.096). This indicates that while the dashboard increased perceived task complexity (mental load), it did not substantially alter the amount of effort students invested in the writing process.

### 4.4 ENA Results

To compare the structural differences in conversational patterns between the two groups, we conducted an Epistemic Network Analysis. The results are visualized in the subtracted network plot (Figure 5), which highlights the distinct ways the EG (blue squares) and the CG (red squares) engaged with the GenAI.

The primary difference between the groups is clearly distributed along the horizontal X-axis (MR1). Based on the co-occurrence of conversational codes, we interpret this axis as representing a spectrum of interaction styles. The right side of the axis, where the EG is located, represents a "Reflective and Evaluative" discourse pattern, characterized by connections involving JUDGMENTAL and INDIRECT_REQUEST questions. In contrast, the left side, where the CG is located, represents an "Instrumental and Information-Seeking" pattern, characterized by connections involving INSTRUMENTAL, FEATURE_SPECIFICATION, and CONCEPT_COMPLETION questions.

A Mann-Whitney U test confirms this visual separation, showing a statistically significant difference between the groups' positions on the X-axis (U = 38.50, *p* < 0.001, r = 0.83), but not on the Y-axis (U = 228.50, *p* = 0.94). This indicates that the two groups structured their interactions in fundamentally different ways along this primary dimension.

Examining the specific network connections in more detail reveals the drivers of this difference. As shown in Figure 5, the EG's discourse was defined by stronger connections between INDIRECT_REQUEST and JUDGMENTAL, as well as FEATURE_SPECIFICATION and JUDGMENTAL. This pattern suggests that the dashboard encouraged students to critically evaluate the AI's output and seek nuanced, high-level feedback. For instance, they were more likely to challenge the AI's suggestions or ask for qualitative assessments of their writing. Conversely, the CG's interaction pattern was characterized by stronger links between FEATURE_SPECIFICATION and INSTRUMENTAL, and between CONCEPT_COMPLETION and VERIFICATION. This indicates a more utilitarian approach, where students without the dashboard support focused on asking for direct instructions, definitions, and simple confirmations to complete the task. Their goal was primarily to get the right information or procedure from the AI, rather than to engage in a collaborative, evaluative dialogue.

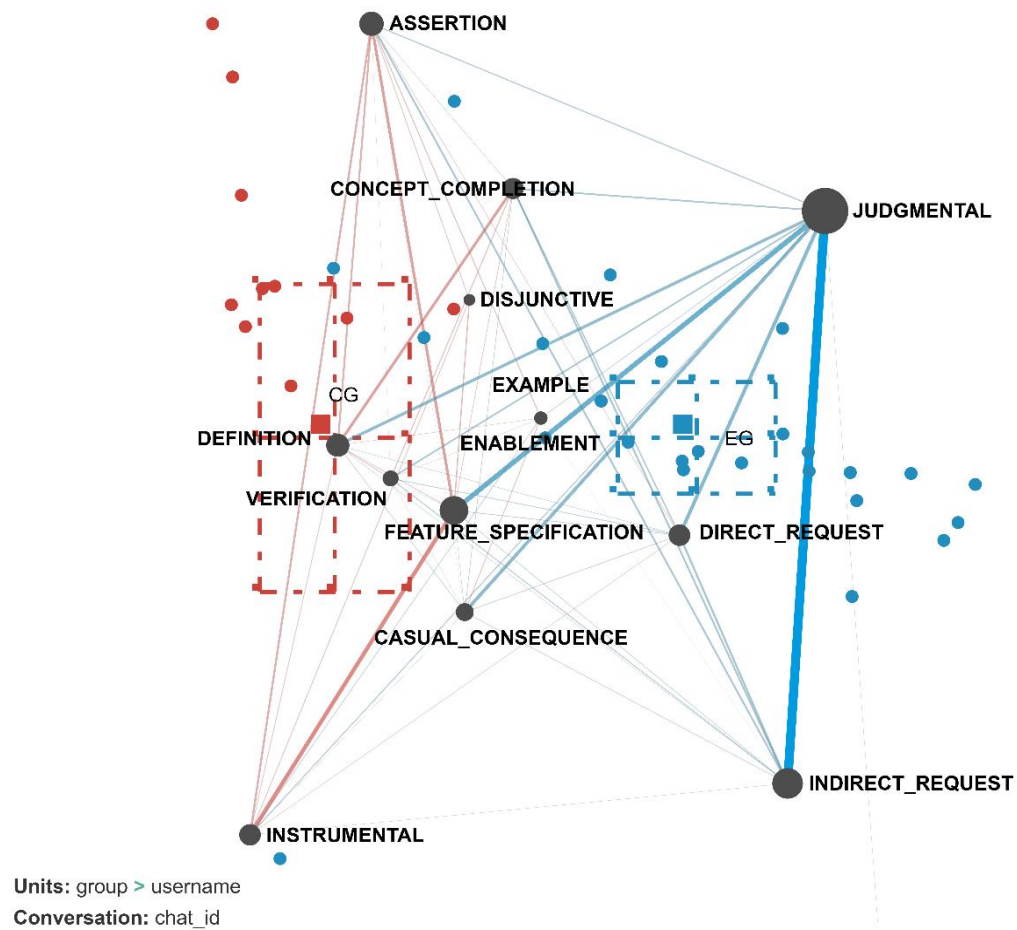

Figure 5 ENA plots of human-AI question-asking between control group and experimental group

# 5. Discussion

## 5.1 Effect of learning analytics dashboard on learning gains in human-AI collaborative writing environment

The results of this study demonstrate that the integration of a LAD in a human-AI collaborative writing environment gave rise to significant improvements in students' academic writing knowledge. We argue that this outcome is not attributable to the writing practice alone, but rather to the dashboard's function as a powerful cognitive scaffold. The mechanism for this knowledge gain can be understood through the specific functions of the dashboard. First, the Abstract Writing Task Preparation Module prompted students to set explicit goals across five critical dimensions of writing quality (e.g., content understanding, structure completeness, logical coherence). This initial step moved students beyond a vague intention to "write a good abstract" and forced them to engage with the specific criteria that define quality, effectively activating their prior knowledge and establishing a clear mental framework for the task. Second, during the writing process, the Writing Progress Monitoring and Writing Reflection Modules provided continuous, formative feedback. By displaying real-time scores on the completeness of abstract components (e.g., background, methods, results) and overall quality dimensions, the dashboard operationalized academic writing standards. This constant, targeted feedback loop transforms abstract rules into practical, in-the-moment problems to be solved. This process reinforces conceptual understanding not merely through dialogue with GenAI, but through direct application and self-evaluation guided by the dashboard.

This finding is particularly significant given the growing concerns about the potential for GenAI tools to foster over-reliance and superficial engagement (Bastani et al., 2024). By providing process-oriented feedback, the dashboard encouraged deeper cognitive engagement in students' learning task. Our findings thus highlight the critical need for structured scaffoldings in human-AI collaborative learning contexts to mitigate the risks of cognitive offloading and ensure that these powerful tools contribute to, rather than undermine, critical thinking and learning gains (Gerlich, 2025).

## 5.2 Effect of learning analytics dashboard on SRL skills and cognitive load in human-AI collaborative writing environment

The analysis of post-intervention self-reported SRL skills and cognitive load revealed a dual-edged effect of the learning analytics dashboard on learners. First, the integration of theory-driven dashboards into learning environments has shown promising effects on SRL skills, particularly in self-efficacy and cognitive strategies. The study by Paulsen and Lindsay (2024) highlights a critical gap in conventional LADs, which often focus on descriptive analytics and outcome-oriented feedback, thereby neglecting the process-oriented aspects of learning. This oversight can limit the potential of LADs to support the dynamic and adaptive nature of SRL, especially in human-AI collaborative writing tasks where learners must balance their own cognitive efforts with the capabilities of GenAI (Nguyen et al., 2024). Compared to descriptive LADs, the theory-driven LAD in this study emphasizes process-oriented feedback grounded in Zimmerman's (2001) SRL framework. The observed enhancements in SRL skills in students' self-reported self-efficacy and their reported use of cognitive strategies, could be attributed to the LAD's ability to support the cyclical phases of SRL—forethought (Writing Preparation Module), performance control (Writing Process Monitoring Module), and self-reflection (Writing Reflection Module). This approach aligns with the growing recognition that effective human-AI collaboration requires learners to engage in dynamic, adaptive learning processes rather than passively relying on AI outputs (Cukurova, 2024; Dellermann et al., 2019; Molenaar, 2022a). For instance, while traditional descriptive LADs often fail to address the process-oriented aspects of learning, the current LAD supports learners in goal-setting, monitoring, and reflection, enabling them to critically evaluate their interactions with GenAI and adjust their strategies accordingly. This may particularly significant in human-AI collaborative writing, where learners must balance their own cognitive efforts with the capabilities of GenAI to avoid over-reliance and superficial learning (Gerlich, 2025; Nguyen et al., 2024). The observed improvements in self-efficacy and cognitive strategies further validate the importance of actionable, process-oriented feedback in fostering self-regulation, as previously highlighted by Duan et al. (2022) and Sedrakyan et al. (2020).

Second, however, the dashboard also introduced some unintended consequences. Notably, the experimental group reported higher levels of test anxiety and cognitive load compared to the control group. Unlike the simple chat interfaces used in many prior GenAI studies (e.g., Noy & Zhang, 2023; Urban et al., 2024), our theory-driven LAD was designed to scaffold and intensify the self-regulatory process. The continuous, real-time feedback from LAD modules likely prompted students to constantly self-evaluate their progress and strategies. This heightened state of self-monitoring, while beneficial for learning gains as shown in our results, can also increase perceived performance pressure and anxiety, as students become more acutely aware of discrepancies between their performance and their goals (Echeverria et al., 2025). Furthermore, the increased cognitive load reported by the experimental group is likely a direct consequence of interacting with the complex, multi-component LAD. According to Cognitive Load Theory (Sweller, 1994), learners were required to simultaneously process multifaceted feedback from the dashboard, integrate this information with GenAI-generated content, and regulate their ongoing writing task. Managing these multiple, interdependent cognitive activities inevitably imposes a higher load. Therefore, the observed increase in test anxiety and cognitive load may be the byproduct of an intervention designed to foster deeper, more reflective learning (Molenaar, 2022a). This aligns with the concept of germane cognitive load, where cognitive load is productively directed towards schema construction and learning (Schnotz & Kürschner, 2007). It highlights a critical design trade-off in human-AI collaborative systems: promoting the effortful cognitive processes essential for meaningful learning may come at the cost of increased perceived task difficulty and psychological pressure.

## 5.3 Effect of dashboard intervention on learner-GenAI discourse in the human-AI collaborative writing environment

The Epistemic Network Analysis (ENA) findings demonstrate that the learning analytics dashboard meaningfully shaped learner-GenAI discourse patterns. The experimental group (EG) exhibited significantly more reflective and evaluative interactions, whereas the control group (CG) primarily engaged in shallow, information-seeking exchanges. This aligns with established research on real-time feedback systems that enhance metacognitive awareness and self-regulated learning in technology-mediated environments (Lim et al., 2023). Three key patterns emerge from the EG discourse: (1) higher frequencies of JUDGMENTAL and INDIRECT_REQUEST question types, (2) stronger associative strength between INDIRECT_REQUEST-JUDGMENTAL codes, and (3) tighter FEATURE_SPECIFICATION-JUDGMENTAL connections. These patterns suggest the dashboard effectively scaffolded critical evaluation of writing outputs and deeper cognitive engagement. For example, one EG participant began their interaction with GenAI by asking, *"Do you think the following passage is coherent?"* (INDIRECT_REQUEST), and later followed up with a JUDGMENTAL question: *"I think this (your answer) is not concise enough, and you added unauthorized parts."* (Details can be found in Appendix 5). Such discourse patterns support previous findings that SRL feedback that can foster learners' capacity to monitor and regulate learning trajectories (Lim et al., 2021).

In contrast, the CG's discourse, characterized by FEATURE_SPECIFICATION and INSTRUMENTAL questions, reflects a more transactional interaction style focused on obtaining procedural guidance. For instance, one CG participant asked, *"How many parts should the abstract's construction be divided into?"* (CONCEPT_COMPLETION), followed by verification questions such as, *"Should I write the conclusion first when writing an abstract?"* (VERIFICATION). (Details can be found in the Appendix 5) This pattern aligns with emerging evidence that learners' unguided use of GenAI prioritizes efficiency over self-regulation (Author, et al., 2024; Lo et al., 2024), potentially fostering tool dependency at the expense of long-term learning outcomes (Gerlich, 2025). The observed difference underscores the dashboard's critical role in mediating human-AI collaboration. By providing the monitoring feedback like AI Dialogue Quality Module, the system appears to scaffold the metacognitive skills necessary for better collaboration with GenAI (Fan et al., 2024). These findings advocate for pedagogical interventions that position GenAI not as an autonomous problem-solver, but as a reflective partner in developing metacognitive skills.

This result also contributes to the growing body of literature on human-AI collaboration in educational settings. Previous studies have shown that learners often struggle to critically evaluate AI-generated content, particularly when the AI's outputs are perceived as authoritative (Kohnke, 2024). The dashboard's effect in fostering evaluative discourse suggests that targeted interventions can mitigate this challenge by encouraging learners to adopt a more critical and reflective approach. This may be particularly relevant in current human-GenAI collaborative writing practice, where the ability to assess and refine AI-generated content is crucial for producing high-quality outputs (Nguyen et al., 2024).

# 6. Implications and Limitations

The findings of this study carry several implications for the research and practice of integrating learning analytics dashboards into human-AI collaborative learning environments. For researchers, these findings emphasize the need to further explore how learning analytics dashboards can be designed to boost human-AI collaboration. This can be achieved by promoting active engagement and critical evaluation of GenAI-generated content (Sedrakyan et al., 2020; Yan, Greiff, et al., 2024). Future research should investigate how dashboards can offer real-time and process-oriented feedback. Such feedback should support metacognitive skills, including self-monitoring and reflection, while minimizing cognitive load and test anxiety (Boehme et al., 2017; Seufert, 2018). Additionally, researchers ought to examine how dashboards can shape the human-AI discourse. Instead of facilitating transactional, information-seeking exchanges, dashboards should encourage reflective and evaluative interactions (Lim et al., 2023; Author, et al., 2024). This involves looking into how dashboards can stimulate judgmental questioning and strategic reflection, which are crucial for long-term learning and knowledge transformation in human-AI collaborative settings (Author, et al., 2024; Cukurova, 2024). Moreover, our study has revealed the critical significance of a solid theoretical foundation in designing learning analytics dashboards. Researchers should endeavor to incorporate established theories, such as self-regulated learning theory, into the design process. By doing so, they can ensure that the dashboards are not only technically sophisticated but also theoretically sound, thus maximizing their potential to support a meaningful learning process.

For educators, when implementing learning analytics dashboards, should take into account the diverse needs and learning styles of students. Since not all students will respond to the dashboard in the same manner, and some may be more sensitive to the increased anxiety and cognitive load it may induce, additional support and training can be provided. For instance, educators could conduct workshops on effectively using the dashboard's feedback, strategies for managing test anxiety, and techniques for reducing cognitive load during the writing process. This personalized approach will ensure that all students can benefit from the dashboard and improve their learning experience in the human-AI collaborative environment. Furthermore, educators can utilize the insights obtained from the dashboard to inform their teaching practices. By analyzing students' interactions with the GenAI tool and the performance data presented on the dashboard, educators can identify areas where students are struggling or excelling. This information can be used to customize the instruction in the human-AI collaborative learning environment, providing targeted support and guidance to students who require it.

However, this study has several limitations. First, the relatively small sample size and homogeneous student population may limit the generalizability of the results. Future research should include students from a wider range of universities and employ a larger sample size. Second, our assessment of SRL relied on specific subscales of the Motivated Strategies for Learning Questionnaire (MSLQ). It is important to acknowledge that MSLQ subscales capture specific facets (e.g., self-efficacy, intrinsic motivation, test anxiety) and not the entirety of the complex SRL construct; for instance, test anxiety represents only one of many emotions experienced during learning. Furthermore, self-report measures for constructs like cognitive strategy use are known to be susceptible to biases and may not always accurately reflect actual deployment, as highlighted in critiques of similar instruments (Chen, et al., 2024; Fan et al., 2021). While informative, these self-reported perceptions should be interpreted with caution, and future research could benefit from triangulating these findings with more objective measures of SRL processes and strategy use. Third, while this pre-post design allows for an examination of changes over the intervention period, it does not capture the dynamic, moment-to-moment influence of the LAD on these psychological constructs during the learning process itself. Thus, we cannot definitively conclude from these measures alone that the LAD directly caused or helped with all observed changes in these self-reported measures as they unfolded during the task. Future studies could benefit from incorporating methods like experience sampling or momentary assessments during the task to provide stronger evidence of the LAD's immediate effects and mediating processes. Fourth, a limitation of this study is our use of a non-standardized, teacher-created test to measure academic writing knowledge. While this test was designed for high content validity within our specific educational context, it lacks the established validity and reliability compared to a standardized instrument. Therefore, although the findings are meaningful for this study, they should be generalized with caution. Future research would be strengthened by using or developing a more rigorously validated knowledge test to confirm these results. Finally, this study's single-session design highlights the need to investigate the long-term effects of the dashboard intervention. While the experimental group demonstrated significant knowledge gains, as well as increased test anxiety and cognitive load, the 130-minute experiment offers only preliminary insights. It remains unclear whether prolonged exposure would sustain the heightened anxiety and load, or if students would adapt over time. Crucially, it is also unknown whether the observed knowledge improvements are durable or represent a short-term effect. Longitudinal studies are therefore essential to track not only the evolution of affective and cognitive load dynamics but also the retention of knowledge gains over multiple writing tasks.

## 7. Conclusion

In conclusion, this study provides empirical evidence that theory-driven learning analytics dashboards can play a critical role in enhancing learning gains and SRL in human-AI collaborative writing environments. By addressing the challenges associated with over-reliance on GenAI tools and promoting active engagement, the dashboard intervention helped students achieve meaningful cognitive improvements. These findings contribute to the growing body of literature on the educational use of GenAI and highlight the importance of designing interventions that complement GenAI tools, ensuring that technology enhances rather than undermines the learning process.

## Declaration

The authors declare that they have no competing interests.

## Declaration of generative AI and AI-assisted technologies in the writing process

During the preparation of this work, the authors used Grammarly and ChatGPT for grammar checks and better readability. After using these tools, the authors reviewed and edited the content as needed and take full responsibility for the content of the publication.

# Appendix

**Appendix 1**

**Explanation of each dimension for learners to set goals:**

Expected time: Total time spent reading comprehension, writing, and revising

Content understanding: The degree of understanding of the core content of the original literature

Structure completeness: The integrity of the abstract structure includes elements such as background, purpose, methods, results, and conclusions

Expression accuracy: The accuracy of academic language expression, including the use of technical terms and language norms.

Logical coherence: The logical coherence of the content in the abstract, with a natural transition between each part.

**Appendix 2**

**Prompts for scoring the completeness of each dimension:**

Research Background: As a professional 250-word academic abstract writing assistant, please analyze whether the user's abstract includes the research background section. Output format: Score, Analysis. 1 indicates no description, 2 indicates very brief description, 3 indicates basic completeness, 4 indicates relatively complete, 5 indicates very complete. Example: 4, The description is relatively complete. The research background is well-presented, providing context and highlighting the significance of the study. However, it could further elaborate on the specific gaps or challenges in the field that the research aims to address.

Research Question: As a professional 250-word academic abstract writing assistant, please analyze whether the user's abstract includes the research question section. Output format: Score, Analysis. 1 indicates no description, 2 indicates very brief description, 3 indicates basic completeness, 4 indicates relatively complete, 5 indicates very complete. Example: 4, The description is relatively complete. The research question is clearly stated, aligning with the study's objectives. However, it could benefit from additional clarity or specificity to fully capture the scope and focus of the research.

Research Method: As a professional 250-word academic abstract writing assistant, please analyze whether the user's abstract includes the research method section. Output format: Score, Analysis. 1 indicates no description, 2 indicates very brief description, 3 indicates basic completeness, 4 indicates relatively complete, 5 indicates very complete. Example: 4, The description is relatively complete. The research method is outlined, providing a general understanding of the approach. However, it lacks some details, such as specific procedures, tools, or data analysis techniques, which would enhance the reader's comprehension of the methodology.

Research Results: As a professional 250-word academic abstract writing assistant, please analyze whether the user's abstract includes the research results section. Output format: Score, Analysis. 1 indicates no description, 2 indicates very brief description, 3 indicates basic completeness, 4 indicates relatively complete, 5 indicates very complete. Example: 4, The description is relatively complete. The results are summarized, highlighting key findings. However, it could provide more specific data or insights to better illustrate the outcomes and their implications for the research question.

Research Conclusion: As a professional 250-word academic abstract writing assistant, please analyze whether the user's abstract includes the research conclusion section. Output format: Score, Analysis. 1 indicates no description, 2 indicates very brief description, 3 indicates basic completeness, 4 indicates relatively complete, 5 indicates very complete. Example: 4, The description is relatively complete. The conclusion effectively summarizes the study's contributions and implications. However, it could further emphasize the broader impact or suggest directions for future research to enhance its depth and relevance.

**Appendix 3**

**Prompts for scoring the actual writing output scores on the corresponding dimensions:**

Logical Coherence: As a professional 250-word academic abstract writing assistant, please evaluate the logical coherence of the user's writing (score 0-100), including transitions between paragraphs, clarity of arguments, and overall structural coherence. Analysis should reference the user's writing. Output format: {"Score":, "Analysis":}. Example: {"Score":85, "Analysis": The arguments are logically clear, and transitions between paragraphs are natural.}

Expression Accuracy: As a professional 250-word academic abstract writing assistant, please evaluate the accuracy of the user's expression (score 0-100), including the use of professional terminology, language standardization, and precision of expression. Analysis should reference the user's writing. Output format: {"Score":, "Analysis":}. Example: {"Score":78, "Analysis": Professional terminology is used appropriately, but some expressions could be more precise.}

Content Understanding: As a professional 250-word academic abstract writing assistant, please evaluate the depth of the user's understanding of the research content (score 0-100), including the grasp of core concepts and the articulation of research significance. Analysis should reference the user's writing. Output format: {"Score":, "Analysis":}. Example: {"Score":82, "Analysis": The core concepts are well understood, and the research significance is clearly articulated.}

Structural Completeness: As a professional 250-word academic abstract writing assistant, please evaluate the structural completeness of the user's writing (score 0-100), including whether all necessary components are included. Analysis should reference the user's writing. Output format: {"Score":, "Analysis":}. Example: {"Score":90, "Analysis": The structure is complete, and the proportions of each section are reasonable.}

**Appendix 4**

**Prompts for scoring the learners' questions to ChatAI:**

Task Focus: As a professional academic abstract writing teaching assistant, please evaluate the user's task focus when seeking help (score 0-100), including whether they consistently stay aligned with the writing goal. Output format: {"Score":, "Analysis":}. Example: {"Score":75, "Analysis": The user generally maintains task focus but occasionally deviates.}

Question Quality: As a professional academic abstract writing teaching assistant, please evaluate the quality of the user's questions when seeking help (score 0-100), including clarity, depth, and relevance. Output format: {"Score":, "Analysis":}. Example: {"Score":85, "Analysis": The questions are clear and demonstrate depth.}

Academic Standards: As a professional academic abstract writing teaching assistant, please evaluate the user's adherence to academic norms when seeking help (score 0-100), including whether they follow academic ethics and avoid over-reliance on AI. Output format: {"Score":, "Analysis":}. Example: {"Score":78, "Analysis": The user generally follows academic norms but has room for improvement.}

Independent Thinking: As a professional academic abstract writing teaching assistant, please evaluate the user's independent thinking ability when seeking help (score 0-100), including their autonomous judgment and critical thinking. Output format: {"Score":, "Analysis":}. Example: {"Score":82, "Analysis": The user demonstrates good independent thinking ability.}

Questioning Strategy: As a professional academic abstract writing teaching assistant, please evaluate the user's questioning strategy when seeking help (score 0-100), including their understanding and effective use of AI suggestions. Output format: {"Score":, "Analysis":}. Example: {"Score":90, "Analysis": The questioning strategy is sound, and the user effectively utilizes AI suggestions.}

**Appendix 5**

**Example of human-GenAI interaction in experimental group:**

For example, one learner begin the conversation with ChatAI about asking hints to revise the written paragraph, and following up with judgmental question that point out where ChatAI provided wrong answers:

Table 1 Example of human-GenAI interaction in experimental group

| Role | Content | Code |
|---|---|---|
| User | Do you think the following passage is coherent? Music is common in daily life and even plays an important role in early childhood development across various cultures. Many studies have explored various aspects of music's impact on people, including its cognitive and emotional effects. However, several research questions in this area remain unexplored. | INDIRECT_REQUEST |
| ChatAI | "The passage you provided is largely coherent, but there are areas where the flow and logical progression can be enhanced. Here's an analysis to help you understand its coherence and potential improvements:<br><br>### Suggestions for Improvement:<br>- **Transition Between Sentences**: …...<br><br>- **Clarification of Contrast**: ……<br><br>### Enhanced Version:<br>- ""Music is common in daily life and even plays an important role in early childhood development across various cultures. Many studies have explored various aspects of music's impact on people, including its cognitive and emotional effects. However, despite this extensive research, several key questions in this area remain unexplored, particularly in the realms of long-term developmental outcomes and cross-cultural variations.""<br>Feel free to ask more specific questions about vocabulary, grammar, or further structural improvements!" | - |
| User | I think this is not concise enough, and you added aspects that were not explored without permission. | JUDGMENTAL |
| ChatAI | Apologies for the oversight. Let's focus on making the passage more concise without adding new aspects. Here's a revised version that aims to maintain the | - |

| | original content while enhancing conciseness: ### Revised for Conciseness: "Music, integral to daily life and early childhood development across cultures, has been widely studied for its cognitive and emotional impacts. Yet, numerous research questions in this field remain unaddressed." …… | |

**Example of human-GenAI interaction in control group:**

For example, one learner begins the conversation with asking ChatAI about detailed concepts and following up with concept verification:

Table 2 Example of human-GenAI interaction in control group

| Role | Content | Code |
|------|---------|------|
| User | how many parts the abstract's construction should be divided into | CONCEPT_COMPLETION |
| ChatAI | "An abstract is typically divided into several key parts to effectively summarize the main components of a research paper or academic article. While the exact structure can vary depending on the field and specific requirements, a common division includes the following four to five parts: …… | - |
| User | should I write the conclusion first when writing an abstract | VERIFICATION |
| ChatAI | Writing the conclusion first in an abstract is not typically recommended. The standard structure of an abstract follows the same sequence as the full paper: Introduction, Methods, Results, and then Conclusion. Here's why this order is beneficial: …… | - |
| User | in the end of the abstract, should I wrire the conclusion again? If not, what should I write | VERIFICATION |
| ChatAI | In the final part of an abstract, you typically do not | |

| | rewrite the conclusion in its entirety. Instead, this section should succinctly summarize the key findings and their implications, and may also include suggestions for future research or practical applications. Here's a breakdown of what to include:<br>…… | |
|---|---|---|

## Appendix 6

**Writing knowledge test**

1. **A good academic abstract should primarily answer which questions?**

    A) Only "What was found?"

    B) "Why was it studied?" and "What was found?"

    C) "Why was it studied?", "How was it studied?", and "What was found?"

    D) Only "How was it studied?"

2. **Which of the following is NOT suitable for inclusion in an abstract?**

    A) The main findings of the research

    B) A brief description of the research methodology

    C) Detailed literature citations and theoretical discussions

    D) A short introduction to the research background

3. **According to the "Professional Writing" criteria, what is the key difference between a 1-point and a 0-point abstract regarding grammatical errors?**

    A) The total number of errors

    B) The type of errors (e.g., spelling vs. grammar)

    C) Whether the errors significantly impede the reader's understanding

    D) The use of inappropriate verb tense

4. **When writing an abstract, which tense is most appropriate for describing the research methodology?**

    A) Present tense

    B) Past tense

    C) Present perfect tense

    D) Future tense

5. **The findings section of an abstract should:**

   A) Only list raw data

   B) Provide specific key findings

   C) Explain all findings in detail

   D) Cite the results of other studies

6. **Which of the following is NOT a primary function of an abstract?**

   A) To help readers quickly understand the research content

   B) To explain the theoretical framework in detail

   C) To summarize the main findings of the study

   D) To highlight the importance of the research

7. **An abstract's "Purpose" section includes the sentence: "Additionally, we spent a great deal of time calibrating our equipment, which was very challenging." According to the rubric, why is this statement problematic?**

   A) It should be in the "Findings" section.

   B) It is a detail for the "Methodological Approach."

   C) It contains irrelevant or unimportant information.

   D) It belongs in the "Contribution to Discipline" section.

8. **If the "Findings" section of an abstract describes results that are not directly related to the research question, how would it likely be scored according to the rubric?**

   A) 3 points

   B) 2 points

   C) 1 point

   D) 0 points

9. **Which of the following should NOT appear in an abstract?**

   A) The research question

   B) Figures and tables

   C) The main findings

   D) The significance of the research

10. **When writing an abstract, which of the following practices is correct?**

    A) Using extensive jargon and abbreviations without definition

B) Including a detailed data analysis process

    C) Using clear and concise language

    D) Adding personal opinions and evaluations

11. **What is the most important function of the introductory statement in an abstract?**

    A) To list all research data in detail

    B) To state the conclusion immediately

    C) To engage the reader and indicate the background and motivation for the study

    D) To define all technical terms

12. **According to the rubric, what should a high-quality "Contribution to Discipline" section articulate?**

    A) How difficult the research process was

    B) How the research advances knowledge or its practical applications

    C) The author's detailed future research plans

    D) A comparison with all other related studies

13. **What is the most professional way to handle an acronym when it is first mentioned in an abstract?**

    A) Use it directly, as readers are expected to know it

    B) Provide the full term, followed by the acronym in parentheses

    C) Use only the full term and avoid the acronym

    D) List all acronyms and their definitions at the end of the abstract

14. **According to the rubric, what is a common flaw in a 2-point "Findings" section?**

    A) The findings are completely missing.

    B) The findings are clear and connected to the purpose.

    C) The findings are presented, but might be unclear or have some information missing.

    D) The findings are unrelated to the scholarship.

15. **The sentence "This paper aims to evaluate the impact of two different teaching methods on student learning efficiency" best fits in which section of the abstract?**

    A) Findings

    B) Methodological Approach

    C) Contribution to Discipline

D) Purpose

16. **When describing the future implications or potential impact of the research, which tense does the rubric suggest can be used?**

    A) Past tense

    B) Present tense

    C) Future tense

    D) Past perfect tense

17. **The writing style of an abstract should be:**

    A) Full of personal emotion and persuasive

    B) Objective, precise, and professional

    C) Colloquial and easy for a general audience to understand

    D) Complex and lengthy to demonstrate academic depth

18. **According to this specific rubric, what is the ideal length for the abstract?**

    A) 100-150 words

    B) 150-200 words

    C) 250-300 words

    D) 400-500 words

19. **What is the most logical order for the components of an abstract?**

    A) Findings -> Methods -> Purpose -> Background

    B) Background & Purpose -> Methods -> Findings -> Contribution & Implications

    C) Methods -> Background -> Contribution -> Findings

    D) Contribution -> Purpose -> Findings -> Methods

20. **A clear "Methodological Approach" section should primarily:**

    A) List the model numbers of all lab equipment

    B) Identify the method used and connect it to the research purpose

    C) Provide a detailed comparison with other possible methods

    D) Hide methodological details to encourage readers to read the full paper

**Appendix 7**

**Cognitive Load questionnaire – Pre-test**

**Mental Load**

1. I anticipate that this learning activity, writing an abstract with human-AI collaboration, will be very difficult for me.
2. I expect that I will find it very difficult to understand the content of the article.
3. I believe the background knowledge required for writing this abstract (e.g., structure, key elements, academic writing conventions) will be very difficult.
4. I predict that the learning process for completing the abstract with human-AI collaboration will be very difficult.

**Mental Effort**

1. I invested a lot of mental effort in this abstract writing activity.
2. I exerted a lot of effort during this abstract writing activity.
3. I felt under a lot of time pressure during this abstract writing activity.
4. I felt very tense during this abstract writing activity.

**Cognitive Load questionnaire – Post-test**

**Mental Load**

5. This learning activity, writing an abstract with human-AI collaboration, was very difficult for me.
6. I found it very difficult to understand the content of the article.
7. I found the background knowledge related to writing this abstract (e.g., structure, key elements, academic writing conventions) very difficult.
8. I found the learning process for completing the abstract with human-AI collaboration very difficult.

**Mental Effort**

5. I invested a lot of mental effort in this abstract writing activity.
6. I exerted a lot of effort during this abstract writing activity.
7. I felt under a lot of time pressure during this abstract writing activity.
8. I felt very tense during this abstract writing activity.